\documentclass[conference]{IEEEtran}
\IEEEoverridecommandlockouts
\usepackage{cite}
\usepackage{amsmath,amssymb,amsfonts}
\usepackage{graphicx}
\usepackage{textcomp}
\usepackage{xcolor}
\usepackage[numbers,sort&compress]{natbib}
\usepackage{multirow}
\usepackage{multicol}
\usepackage{booktabs}

\usepackage{comment}
\usepackage{etoolbox}

\newbool{inccomment}
\setbool{inccomment}{true}
\newcommand{\hl}[1]{\ifbool{inccomment}{{\color{red}#1}}{}}
\newcommand{\bl}[1]{\ifbool{inccomment}{{\color{blue}BL: #1}}{}}
\def\BibTeX{{\rm B\kern-.05em{\sc i\kern-.025em b}\kern-.08em
    T\kern-.1667em\lower.7ex\hbox{E}\kern-.125emX}}
\begin{document}

\title{RED: A ReRAM-based Deconvolution Accelerator}
\vspace{-48pt}
\author{\IEEEauthorblockN{Zichen Fan\IEEEauthorrefmark{1}$^\S$,
Ziru Li\IEEEauthorrefmark{1}$^\S$,
Bing Li\IEEEauthorrefmark{1}$^\dagger$$^\ddagger$, 
Yiran Chen$^\dagger$,
Hai (Helen) Li$^\dagger$
}
\IEEEauthorblockA{$^\S$ECE Dept., Tsinghua University, Beijing, China}
\IEEEauthorblockA{$^\dagger$ECE Dept., Duke University, Durham, NC, USA  $^\ddagger$Army Research Office, Research Triangle Park, USA}
\IEEEauthorblockA{$^\S$ \{fanzc15, lizr15\}@mails.tsinghua.edu.cn$ ^\dagger$\{bing.li.ece, yiran.chen, hai.li\}@duke.edu}
\thanks{\IEEEauthorrefmark{1}These three authors contributed equally to this work.}
\thanks{$^\ddagger$This author is supported by the NRC Associate Fellowship Award and is the corresponding author. bing.li.ece@duke.edu}
}


\maketitle

\begin{abstract}

Deconvolution has been widespread in neural networks.
For example, it is essential for performing unsupervised learning in generative adversarial networks or constructing fully convolutional networks for semantic segmentation. 
Resistive RAM (ReRAM)-based processing-in-memory architecture has been widely explored in accelerating convolutional computation and demonstrates good performance. 
Performing deconvolution on existing ReRAM-based accelerator designs, however, suffers from long latency and high energy consumption because deconvolutional computation includes not only convolution but also extra add-on operations. 
To realize the more efficient execution for deconvolution, we analyze its computation requirement and propose a ReRAM-based accelerator design, namely, RED. 
More specific, RED integrates two orthogonal methods, the \textit{pixel-wise mapping scheme} for reducing redundancy caused by zero-inserting operations and the \textit{zero-skipping data flow} for increasing the computation parallelism and therefore improving performance. 
Experimental evaluations show that compared to the state-of-the-art ReRAM-based accelerator, RED can speed up operation 3.69$\sim$31.15$\times$ and reduce 8\%$\sim$88.36\% energy consumption.
\end{abstract}

\section{Introduction}
\label{sec:intro}
Generative adversarial networks (GANs) and fully convolutional networks (FCNs) have been widely explored for their superior performance in processing complicated image tasks. 
For example, GANs are used to reconstruct 3D models from 2D images~\cite{wu2016learning} and recover corrupted images~\cite{yeh2016semantic}. 
FCNs are applied in semantic segmentation~\cite{FCN} and object detection~\cite{zhang2018single}. 
\textit{Deconvolution} layers are the important for these networks to carry out the up-sampling from low-resolution to high-resolution images. 
As most of the current platforms are mainly optimized for the regular convolution, the deconvolutional computation suffers from low-efficiency due to the involved additional operations.
Designing an efficient accelerator catering to deconvolutional computation is considerable significant and taken as the focus of this work. 

Among the existing neural network accelerator, processing-in-memory (PIM), which moves the computation close to and even within the memory elements, demonstrates great potentials. 
Resistive RAM (ReRAM) has been taken as a competitive technology for PIM implementation due to its low-energy and high-efficient vector-matrix multiplications performed on the crossbar structure. 
Various ReRAM-based accelerators ~\cite{chi2016prime,cheng2018time,shafiee2016isaac,song2017pipelayer,qiao2018atomlayer,li2018reram} have been presented for fast and efficient convolution, showing great advantages of ReRAM over the CMOS-based counterparts.

Unfortunately, the unique computation patterns of deconvolution make its implementation on existing ReRAM-based accelerators very challenging.
For example, the common \textit{zero-padding} in deconvolution inserts plenty of zeros into input feature maps before convolution, resulting in massive redundant operations.
The \textit{padding-free} deconvolution excludes the zero-inserting but involves extra operations, \textit{i.e.}, addition and cropping after convolution. 
Though \textit{padding-free} is more friendly for the CMOS-based accelerators~\cite{FCNEngine}, the add-on operations leads to the modified circuits on ReRAM-based accelerator. 

This work aims to develop an efficient ReRAM-based deconvolution accelerator. 
In the work, we first analyze the efficiency of zero-padding and padding-free deconvolution algorithms on existing ReRAM-based platforms. 
Considering the inefficiency in performing zero-padding and the high overhead induced by padding-free, we propose RED, a ReRAM-based accelerator tailored for deconvolutional computation. 
Our approach integrates the optimization on data mapping and data flow. 
More specific, the \textit{pixel-wise mapping} can dramatically reduce the redundancy caused by zero-inserting operations, and the \textit{zero-skipping data flow} further elevates the computation parallelism without add-on periphery circuits.

We evaluated the power, latency, and area of RED when performing the deconvolutional layers in GANs and FCNs and compared with state-of-the-art ReRAM-based accelerators of zero-padding design and padding-free design~\cite{ReGAN}. 
Experimental results show that RED achieve 3.69$\sim$31.15$\times$ speedup and 8\%$\sim$88.36\% energy consumption reduction, with 22.14\% increment in design area. 

This paper is organized as follows. 
Section~\ref{sec:preliminary} introduces the background knowledge including ReRAM-based accelerator and the deconvolutional computation.
Section~\ref{sec:accelerator} elaborates the principle and implementation of RED.
In Section~\ref{sec:experiment}, we evaluate RED in terms of power, latency and area and compare it with state-of-the-art ReRAM-based counterparts.
At the end, we conclude the paper in Section V.
\section{Preliminary}
\label{sec:preliminary}

\subsection{ReRAM-based CNN Accelerator}
\begin{figure}[t]
\begin{center}
\includegraphics[width=1\columnwidth]{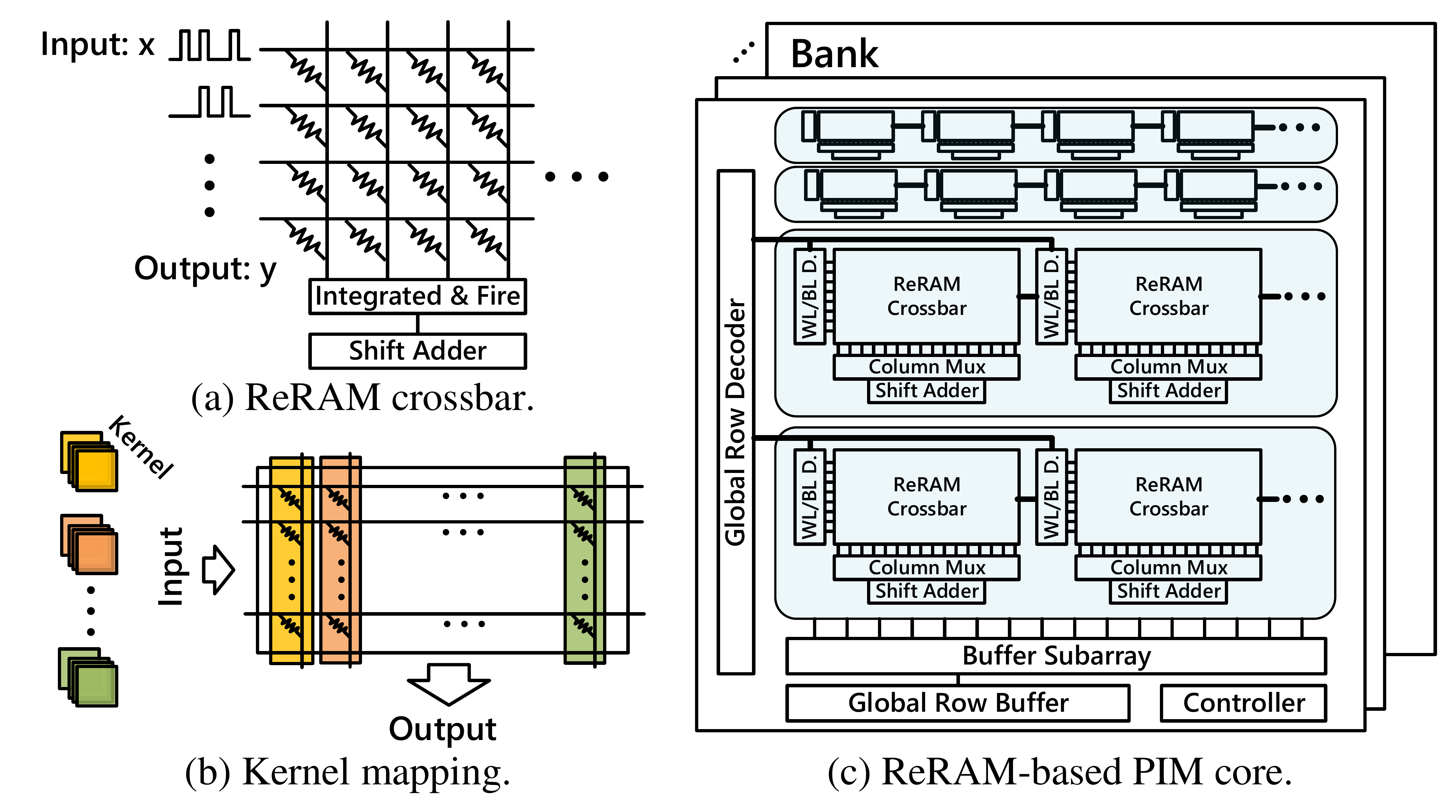}
\vspace{-18pt}
\caption{ReRAM crossbar, kernel mapping and ReRAM-based PIM.}
\label{rram}
\end{center}
\vspace{-6pt}
\end{figure}

As an emerging memory technology, ReRAM crossbar structure can also effectively execute vector-matrix multiplication operations, which have gained significant attention.
As illustrated in Fig.~\ref{rram}(a), the elements of a weight matrix are represented as the conductance of ReRAM cells located at the cross-point of the wordlines and bitlines.
During the operation, an input vector in the form of voltage spikes enters the crossbar along the wordlines, a.k.a., rows, and the currents flowing out from the bitlines, a.k.a., columns, denote the computed output vector of the vector-matrix multiplication.
The integrated \& fire circuit converts the output currents to the digital output data, which is then summed up together by the shift adder.

By leveraging the ReRAM structure, various CNN accelerators have been proposed for the inference or training~\cite{chi2016prime,song2017pipelayer}.
%
The kernel in a convolution layer is a tensor with 4 dimensions and its mapping on crossbar requires a complicated design~\cite{song2017pipelayer,qiao2018atomlayer}.
Fig.~\ref{rram}(b) illustrates an example of the kernel mapping design in the ReRAM-based accelerator. 
The filters of $C$ channels spread into a one-dimension vector and is stored in one column.

Fig.~\ref{rram}(c) depicts a full ReRAM-based PIM architecture~\cite{chi2016prime,song2017pipelayer}, which is based on the main memory structure with supportive periphery circuits.
For instance, the wordline/bitline drivers (\texttt{WL/BL D.} in Fig.~\ref{rram}) generates input pulses and controls the switch of each cell.

\subsection{Deconvolutional Computation}
\label{subsec:deconv}
\begin{figure}[t]
\begin{center}
\includegraphics[width=1\columnwidth]{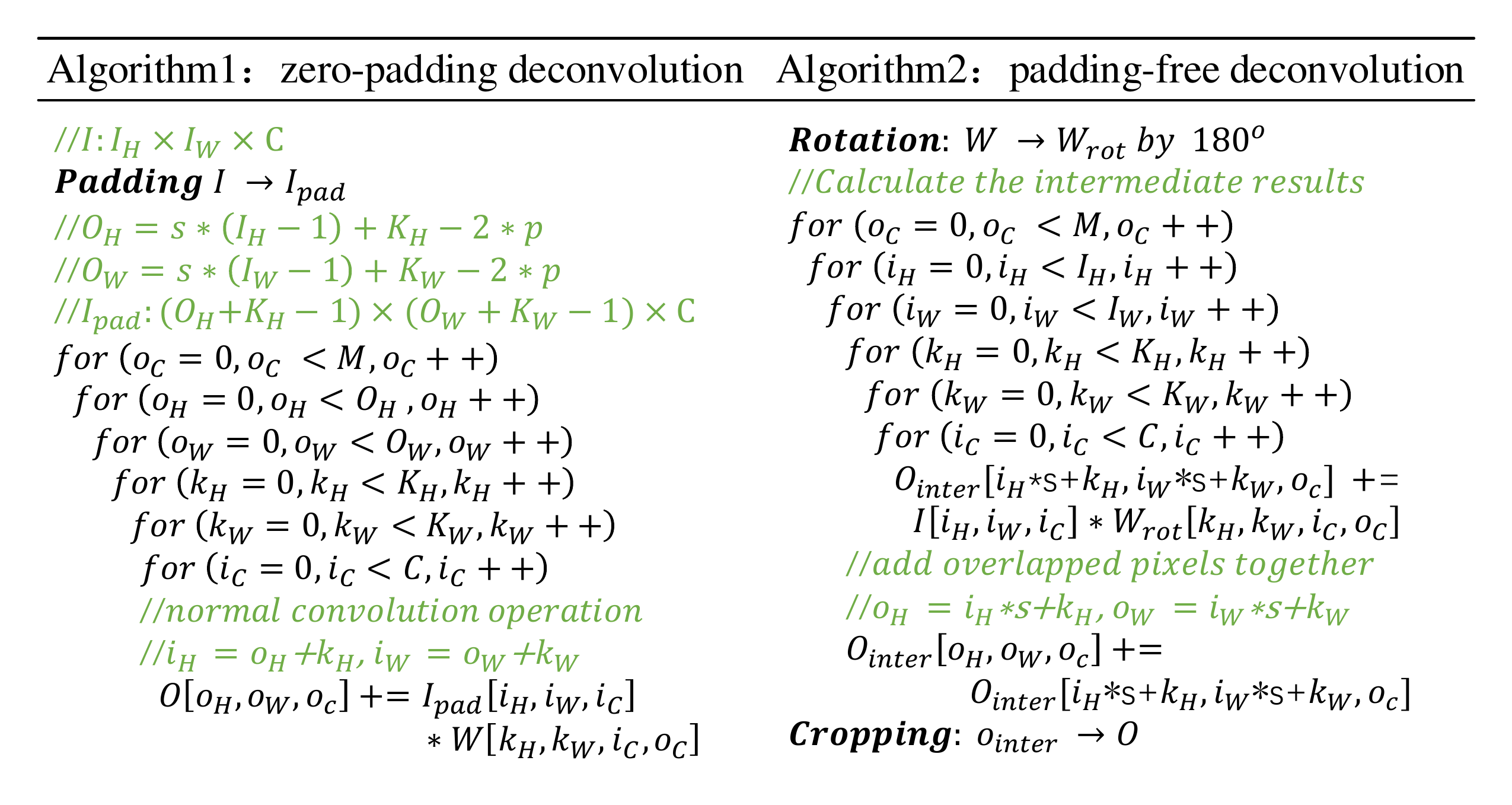}
\vspace{-6pt}
\caption{Pseudo codes of traditional deconvolution algorithms.}
\label{fig:algorithm}
\end{center}
\vspace{-6pt}
\end{figure}

%
Fig.~\ref{fig:algorithm} illustrates two kinds of deconvolutional algorithms: \textit{zero-padding} and \textit{padding-free}.
Similar to convolutional computation, stride $s$ and padding $p$ are two hyper-parameters in the deconvolutional computation.
Suppose that the input data $I$ and consists of a serial feature maps is a $I_H \times I_W \times C$ tensor. Here, $I_H$ and $I_W$ are respectively the height and width of each input feature map. $C$ is the number of channel. The convolution kernel $K$ contains a set of filters and is represented by a $K_H \times K_W \times C \times M$ tensor. $M$ is the number of filters, which is equivalent to the number of output feature maps. Like the input data, the output $O$ composes of $M$ feature maps each of which is $O_H \times O_W$. In this work, each element in the input and output is referred as \textit{pixel}. 
The deconvolution is a up-sampling operation and therefore $O_H \geq I_H$ and $O_W \geq I_W$. 


The zero-padding deconvolution (Algorithm 1) includes two steps:
a) Padding: Insert zeros between the pixels in the input feature maps (denoted as $I_{pad}$);
b) Convolution: Perform regular convolution for $I_{pad}$ with kernels.
As can be seen, zero-padding has massive redundant multiplications as the zero-value input operands in the padding input feature maps. 

Algorithm 2 describes the padding-free algorithm with the following four major steps:
a) Rotation: rotate the weight kernel by $180^\circ$;
b) Convolution: compute the intermediate results by multiply-and-accumulating (MAC) an input pixel with the corresponding kernel in the channel direction;
c) Addition: add the overlapped pixels obtained in step b) together; and
d) Cropping: crop the data at the edge of the output matrices to fit the size of the final output.
Padding-free algorithm avoids inserting zero into the input in comparison with zero-padding.
However, it introduces two additional operations---addition and cropping.
Previously, Xu~\textit{et al.}~\cite{FCNEngine} successfully utilize the padding-free algorithm to adapt the CMOS-based hardware for efficient deconvolutional computation. 
As we shall show in Section~\ref{subsec:motivation}, the existing ReRAM-based accelerators need substantial efforts to realize these operations, incurring a large overhead.

\section{ReRAM-based Deconvolution Accelerator}
\label{sec:accelerator}

In this section, we first analyze the computation inefficiency when mapping the two popular deconvolutional algorithms to the existing ReRAM-based accelerators.
Then we elaborate the proposed RED: a ReRAM-based deconvolution accelerator design which exploits \textit{pixel-wise mapping} and \textit{zero-skipping data flow} to perform high-efficient deconvolution computation.
We also analyze the tradeoff in RED between the area overhead and parallelism.

\subsection{Analysis \& Observations}
\label{subsec:motivation}

\begin{figure}[t]
\centering
\includegraphics[width=0.9\columnwidth]{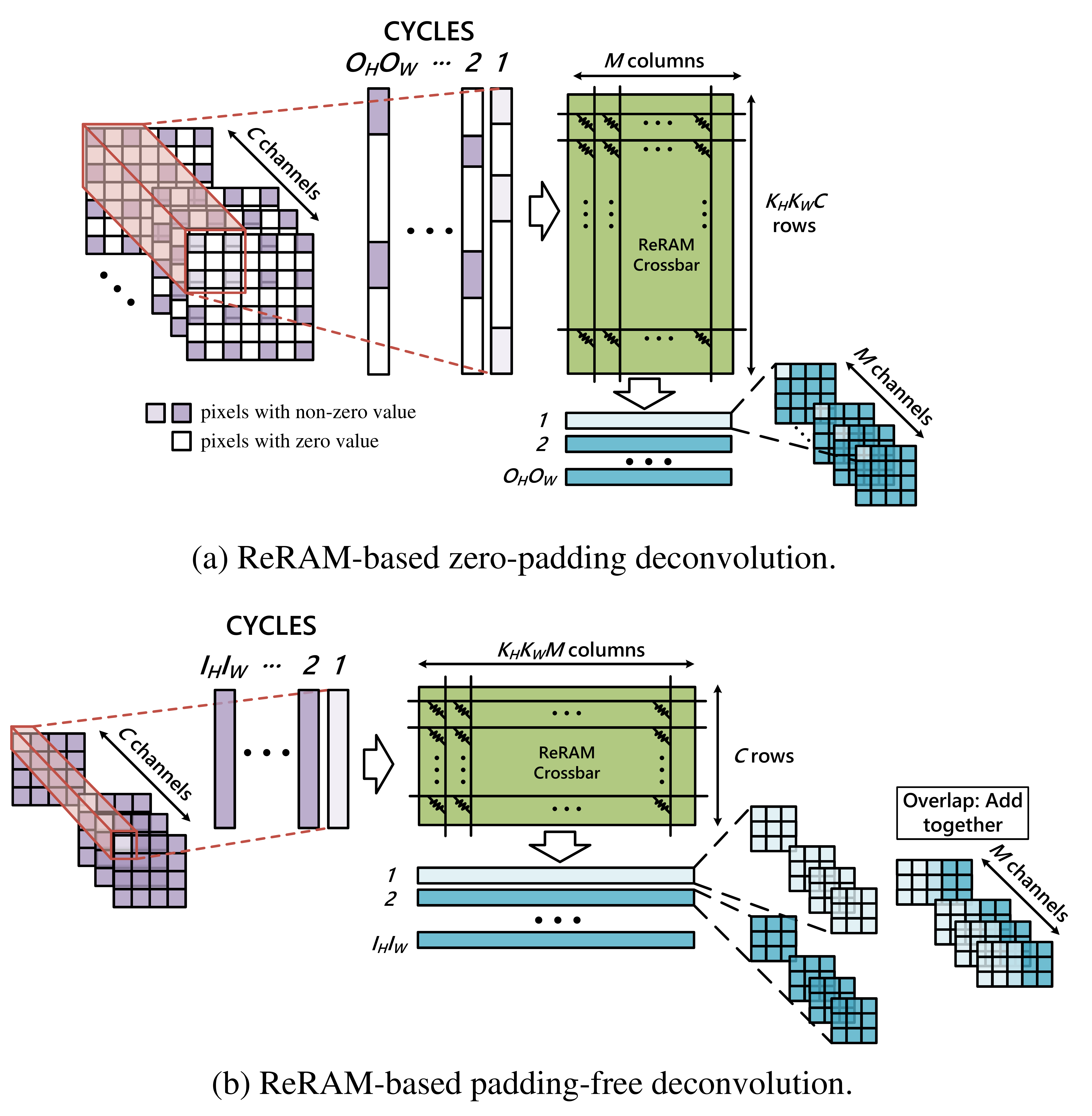}
\vspace{-6pt}
\caption{Deconvolution on ReRAM-based accelerator.}
\label{paddingandpaddingfree}
\vspace{-18pt}
\end{figure}
\begin{figure}[t]
\begin{center}
\includegraphics[width=0.75\columnwidth]{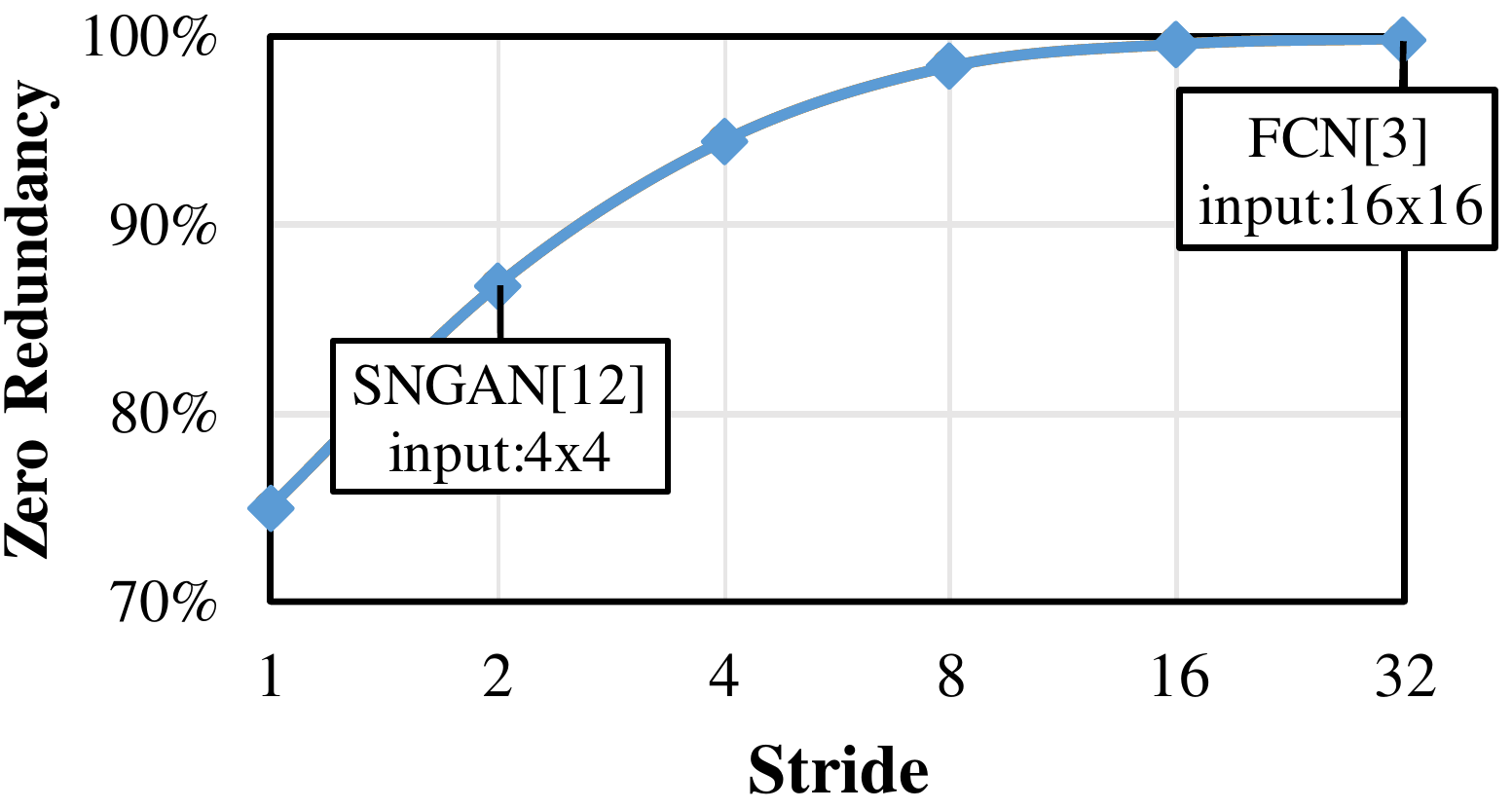}
\vspace{-12pt}
\caption{The zero redundancy ratio in zero-padding deconvolution changing with the stride.}
\label{zeroredundancy}
\end{center}
\vspace{-12pt}
\end{figure}

Fig.~\ref{paddingandpaddingfree}(a) illustrates the zero-padding deconvolution implementation.
The kernel mapping of zero-padding deconvolution is the same as the standard convolutional computation described in Section~\ref{sec:preliminary}:
$M$ weights in deconvolutional layer are mapped on $M$ columns of a ReRAM crossbar. 
In each cycle, one input vector is fed into the crossbar for computation and each pixel in the produced $M$-bit output vector corresponds to one-pixel information for $M$ output feature maps.
As such, it will take $O_H \times O_W$ cycles to obtain the completed data of the $M$ output feature maps in the shape of $O_H \times O_W$.
After the padding step (Section~\ref{subsec:deconv}), the input vector has inserted a large number of zeros and becomes very sparse, inducing the redundant computations on the zero pixel. 
Fig.~\ref{zeroredundancy} presents the zero redundancy ratio (\textit{i.e.}, the ratio of redundant computation induced by zero-padding over total computation) when varying the stride. 
Typically, the deconvolution layer in GANs (\textit{e.g.}, SNGAN~\cite{SNGAN}) sets the stride step to 2, while FCNs~\cite{FCN} usually prefer larger strides in deconvolution layers, such as 8, 16, or 32. 
As shown in Fig.~\ref{zeroredundancy}, the zero redundancy ratio is already $86.8\%$ when $\mathrm{stride}=2$ and grows up to amazingly $99.8\%$ when $\mathrm{stride}=32$. 
The high zero redundancy ratios indicates there are a large amount of redundant operations when ReRAM-based accelerator performs deconvolution.
Note that ReGAN~\cite{ReGAN} adopted the zero-padding deconvolution but neglected the redundant operations. 

Padding-free is an alternative deconvolution algorithm that escapes the zero redundancy.
The previous study~\cite{FCNEngine} showed that the padding-free deconvolution achieved up to 44.9$\times$ performance improvement on the CMOS-based platforms, such as ASIC.
However, our analysis shows that direct mapping the padding-free algorithm on a ReRAM architecture might not be efficient.
As depicted in Fig.~\ref{paddingandpaddingfree}(b), different from the zero-padding deconvolution with a compacted output in $M$ columns, the implementation of padding-free deconvolution on a crossbar requires $K_H\times K_W\times M$ columns.
As the wordline/bitline driving power increases in a quadratic relation with the column number, the padding-free deconvolution expects a much higher power consumption than the zero-padding deconvolution.
What's more, the output from the crossbar is not the final result but requires further processing (addition and cropping), which leads to dedicated circuit support and extra area cost. 

\vspace{-6pt}
\subsection{RED Architecture}
\vspace{-3pt}
\begin{figure*}[t]
\centering
\includegraphics[width=1\textwidth]{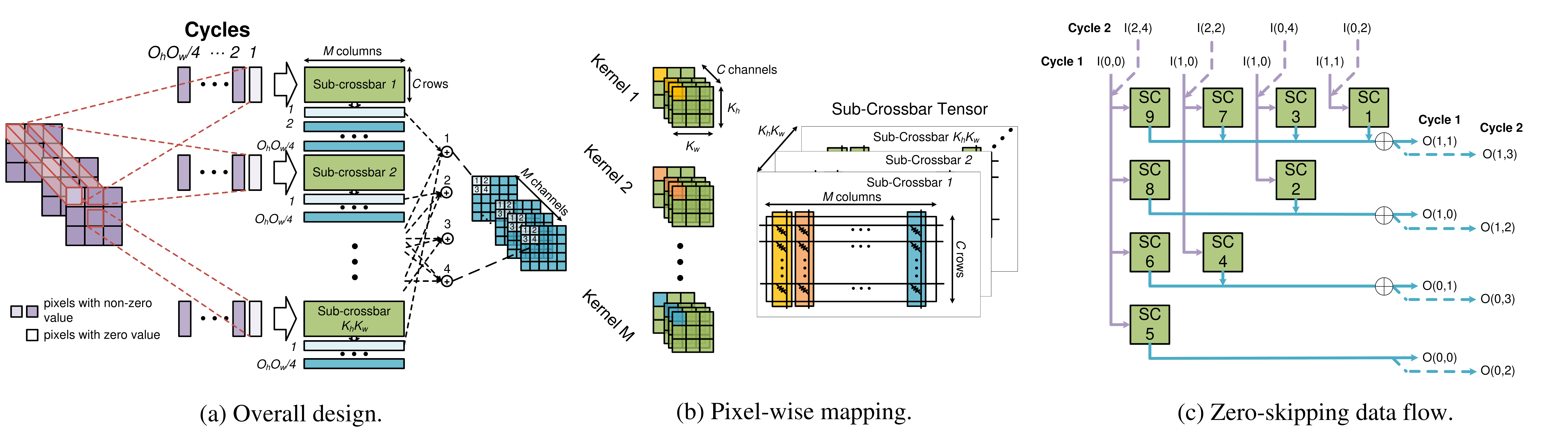}
\vspace{-6pt}
\caption{The illustration of RED architecture(a), pixel-wise mapping(b) and zero-skipping data flow(c).}
\label{overall}
\vspace{-12pt}
\end{figure*} 

\begin{figure}[b]
\begin{center}
\vspace{-6pt}
\includegraphics[width=0.4\textwidth]{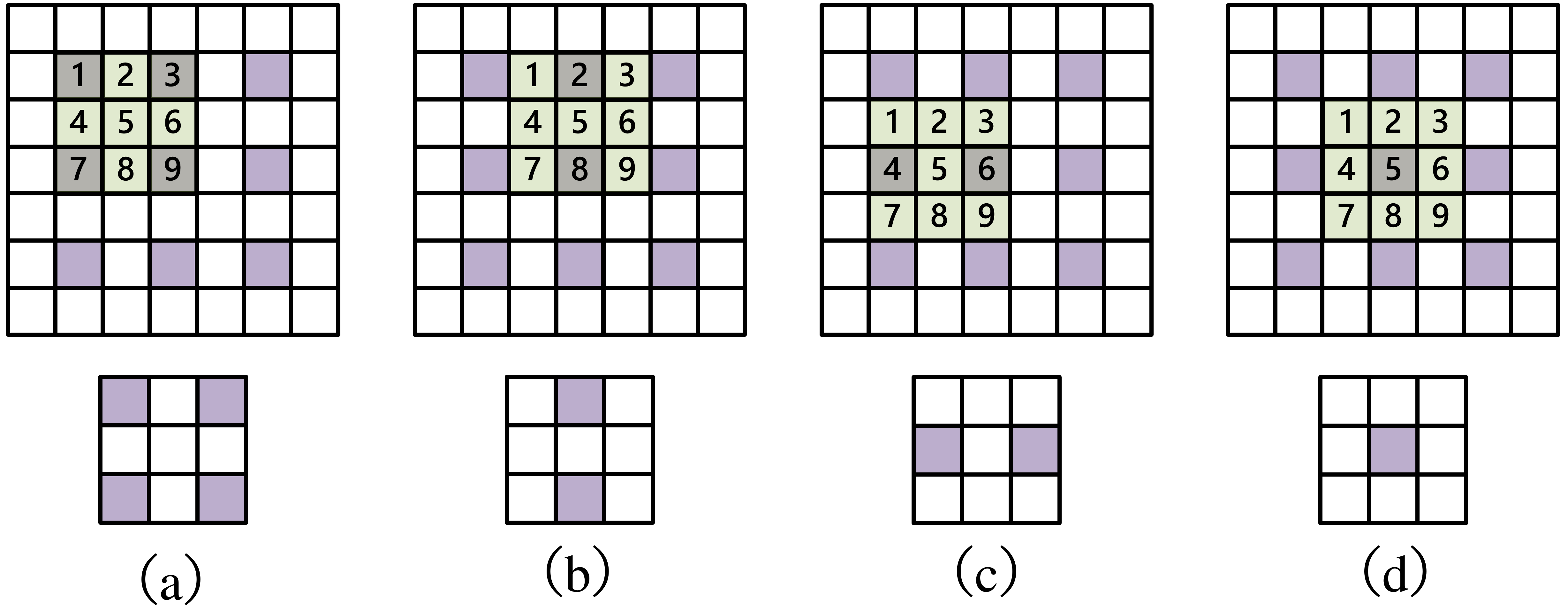}
\vspace{-6pt}
\caption{The four computation modes in deconvolution when the kernel  size is $3\times3$ and the stride is 2.}
\label{4modes}
\end{center}
\end{figure}

To overcome the aforementioned problems, we propose RED---a new ReRAM-based deconvolution accelerator. 
The design combines two orthogonal approaches, respectively for minimizing the redundant operations induced by the padded zeros and for enhancing the execution parallelism without additional operations. 
For ease of the explanation, we take a deconvolutional computation with $\mathrm{stride} = 2$ and the kernel filter size of $3\times 3$ as the example in the following description.

The overall RED architecture is presented in Fig.~\ref{overall}(a).
Here, the computation of the deconvolution is executed by $K_H\times K_W$ sub-crossbars (denoted as ``SC") each of which is the size of $C\times M$.
To clarify, we demonstrate the padded zeros in Fig.~\ref{overall}.
During the deconvolutional computation, RED takes only those non-zero pixels (in purple) to form the input vectors. 
The partial results from the corresponding sub-crossbars are summed up to obtain the output pixels. 
In each clock cycle, multiple pixels for each output feature map are generated concurrently.
In the following, we will elaborate the details of the pixel-wise mapping in Fig.~\ref{overall}(b) and zero-skipping data flow in Fig.~\ref{overall}(c).
%
\subsubsection{Pixel-wise Mapping}
We propose the pixel-wise mapping to eliminate the high zero redundancy induced by zero-padding algorithm.
Fig.~\ref{4modes} explains the design principle.
In the figure, the large grid refers to a padded input feature map, whose non-zero and zero pixels are denoted in purple and white colors, respectively. 
The small grid with numbers indicates the kernel with its weight location, in which only the purple bricks are the valid weights in utilization. 
Due to the high redundancy of the padded image, only a small portion of weights take part in the convolution operation.

Fig.~\ref{4modes}(a)$\sim$(d) illustrates the four computation modes when sliding the kernel filter within the input feature map. 
For the given configuration, a kernel filter has nine weights, labeled with numbers $1\sim 9$. 
Starting from the first convolutional computation in Fig.~\ref{4modes}(a), there are only four weights ($1$, $3$, $7$ and $9$) contributing to the calculation result. 
The following convolution by sliding the kernel filter horizontally one step involves only two weights $4$ and $6$, as shown in Fig.~\ref{4modes}(b). 
Similarly, the computation modes in Fig.~\ref{4modes}(c) \& (d) occur when moving the kernel window down one grid from the positions in Fig.~\ref{4modes}(a) \& (b), respectively. 
%
%
%
We observe that the convolution operations in the deconvolution are the repetition of the four computation modes.
Furthermore, the weights of the kernel filter are exclusive among these modes. 
Thus, we propose \textit{pixel-wise mapping} to execute the computation modes (a)$\sim$(d) in parallel. 

%

We map a kernel in size of $K_H\times K_W\times C\times M$ into $K_H\times K_W$ sub-crossbars. Each sub-crossbar has $C$ inputs and $M$ outputs, thus can be expressed as a matrix whose shape is $C\times M$. Suppose that combining all the sub-crossbars can form a sub-crossbar tensor ($\mathbf{SCT}$), whose shape is $C\times M\times (K_H\times K_W)$, as shown in Fig.~\ref{overall}(b), then our pixel-wise mapping approach can be expressed as:
\begin{equation}
\mathbf{SCT}[c,m,i*K_W+j] = \mathbf{W}[i,j,c,m], 
\end{equation}
where $0\leq i < K_H$ and $0\leq j< K_W$ indicate the location of the weight in each filter, $0\leq c < C$ denotes the $c^{th}$ channel of the weight filter, $0\leq m < M$ refers to $m^{th}$ weight filter. 

Once a round of computation in sub-crossbars is completed, we add the output from corresponding SCs to obtain the final deconvolution results. 
Thanks to the vertical sum-up design in the existing ReRAM-based accelerators~\cite{song2017pipelayer,ReGAN}, no extra circuitry is needed to realize the addition operations in \textit{pixel-wise mapping}.

\subsubsection{Zero-skipping Data Flow}
Based on the pixel-wise mapping scheme, we develop the zero-skipping data flow which takes non-zero pixels as the inputs of $\mathbf{SCT}$. 
Fig.~\ref{overall}(c) illustrates the operation for the given example with $\mathrm{stride}=2$ and kernel size of $3\times 3$.
Accordingly, there are 9 sub-crossbars. 
The output vectors from the sub-crossbars on the same row will be added up together; and sub-crossbars along the same column will take the same input vectors.
For brevity, we use $\mathbf{I}(i,j)$ to denote the input vector and has $C$ pixels each of which from one channel.
The proposed data flow bypasses padded zeros, so $i$ and $j$ corresponding to the index on padded image are always \textit{even} numbers.
In $\mathrm{Cycle~1}$, $\mathbf{I}(0,0)$ goes to \texttt{SC1}, $\mathbf{I}(2,0)$ is provided to \texttt{SC2} and \texttt{SC3}, $\mathbf{I}(0,2)$ is taken by \texttt{SC4} and \texttt{SC7}, and $\mathbf{I}(2,2)$ is applied to \texttt{SC5}, \texttt{SC6}, \texttt{SC8} and \texttt{SC9}. 
The 9 sub-crossbars operate simultaneously and their outputs will be put together upon the above explanation for the final deconvolution results.  
In the following cycle, RED continues to compute the kernels with the next batch of non-zero pixels, \textit{e.g.}, $\mathbf{I}(0,2)$, $\mathbf{I}(0,4)$, $\mathbf{I}(2,2)$ and $\mathbf{I}(2,4)$ in $\mathrm{Cycle~2}$ as illustrated in the figure. 
Compared to the zero-padding deconvolution, the zero-skipping data flow increases the computation parallelism of this example $4\times$.
\begin{table*}[t]
\centering
\footnotesize
\caption{Benchmarks used in this work}
\vspace{-3pt}
\label{benchmark}
\begin{tabular}{l l l l l l l}
\toprule
Layer Name & Network Model & Dataset               & \begin{tabular}[c]{@{}l@{}}Input Size\\ ($I_H,I_W,C$)\end{tabular} & \begin{tabular}[c]{@{}l@{}}Output Size\\ ($O_H,O_W,M$)\end{tabular} & \begin{tabular}[c]{@{}l@{}}Kernel Size\\ ($K_H,K_W,C,M$)\end{tabular} & Stride \\ \midrule
GAN\_Deconv1 & DCGAN~\cite{DCGAN} & LSUN  & (8, 8, 512) & (16, 16, 256)     & (5, 5, 512, 256) & 2 \\ 
GAN\_Deconv2 & Improved GAN~\cite{improvedGAN}  & Cifar-10 & (4, 4, 512) & (8, 8, 256) & (5, 5, 512, 256) & 2\\ 
GAN\_Deconv3 & SNGAN~\cite{SNGAN} & Cifar-10 & (4, 4, 512) & (8, 8, 256) & (4, 4, 512, 256) & 2 \\ 
GAN\_Deconv4 &SNGAN~\cite{SNGAN}& STL-10  & (6, 6, 512) & (12, 12, 256) & (4, 4, 512, 256) & 2 \\ 
FCN\_Deconv1 & voc-fcn8s\_2x~\cite{FCN} & PASCAL VOC & (16, 16, 21) & (34, 34, 21) & (4, 4, 21, 21) & 2 \\ 
FCN\_Deconv2 & voc-fcn8s\_8x~\cite{FCN} & PASCAL VOC & (70, 70, 21) & (568, 568, 21) & (16, 16, 21, 21) & 8 \\ 
\bottomrule
\vspace{-18pt}
\end{tabular}
\end{table*}
\vspace{-3pt}
\subsection{Design Trade-off}
We use $\mathrm{stride}=2$ to illustrate the RED design.
The deconvolution with $\mathrm{stride}=2$ can be decomposed into four computation modes and therefore achieve 4$\times$ speedup by RED.
The number of computation modes is $\mathrm{stride^2}$, indicating the speed-up brought by RED quadratically increases with the stride.

The kernel size usually grows with the stride. 
For the FCN~\cite{FCN} with $\mathrm{stride}=8$, the kernel filter size is $16\times16$.
Accordingly, 256 sub-crossbars are needed to complete the entire computation modes simultaneously.
More sub-crossbars can cause the increment of the area due to the extra wordline/bitline driver, column mux, shift-adder, \textit{etc.}). 

There exists a trade-off between the area and the execution speeup in RED. 
When the kernel filter size is too large, RED can take more time for computation in exchange for the area efficiency. 
We can reduce the number of sub-crossbars to half of its original number by adding zeros to the input vector.
Suppose that the size of sub-crossbar tensor SCT is $ C\times M\times (K_H\times K_W)$.
In area-efficient design, the shape of SCT shape is $2C\times M \times \frac{K_H\times K_W}{2}$. The data flow changes as below:
\begin{equation}
\small
\begin{aligned}
\mathrm{Cycle~1:}&~~I_n[c]_{c=1,...,C} = I_{2n,ori}[c]_{c=1,...,C}; \\
&~~I_n[c]_{c=C+1,...,2C}=0;\\
\mathrm{Cycle~2:}&~~I_n[c]_{c=1,...,C} = 0; \\
&~~I_n[c]_{c=C+1,...,2C}=I_{2n+1,ori}[c]_{c=1,...,C};
\end{aligned}
\end{equation}
where $I_n$ denotes the input vector of $n^{th}$ modified sub-crossbar ($0\leq n < \frac{K_HK_W}{2}$) and $I_{n,ori}$ is the original input vector in the pixel-wise mapping method.
In this way, we employ 128 sub-arrays to complete the 64 computation modes in two cycles when $\mathrm{stride}=8$ and kernel filter size of $16\times16$.

\begin{table}[t]
\centering
\footnotesize
\caption{Breakdown Component}
\vspace{-3pt}
\label{breakdown}
\begin{tabular}{l l l}
\toprule
\multicolumn{1}{l }{} & \textbf{Component} & \textbf{Abbr.} \\ \midrule
\multirow{3}{*}{Array ($a$)} & Computation & $c$ \\ \cmidrule (l){2-3} 
 & Wordline Driving & $wd$ \\ \cmidrule(l){2-3} 
 & Bitine Driving & $bd$ \\ \midrule
\multirow{4}{*}{Periphery ($pp$)} & Multiplexer & $mux$ \\ \cmidrule(l){2-3} 
 & Decoder & $dec$ \\ \cmidrule(l){2-3} 
 & Read Circuit / Integrated \& Fire Circuit & $rc$ \\ \cmidrule(l){2-3} 
 & Shift Adder & $sa$ \\ 
 \bottomrule
\end{tabular}
\vspace{-18pt}
\end{table}

\section{Experiments}
\label{sec:experiment}

This section evaluates RED in terms of performance, energy consumption, and area overhead. 
We compare RED with the conventional zero-padding and padding-free design using the deconvolutional layers from representative neural networks. 

\vspace{-3pt}
\subsection{Experimental Setup}
\vspace{-3pt}
We modified NeuroSim+~\cite{Chen2018NeuroSim} to implement the conventional zero-padding design, padding-free design, and our proposed RED design. 
The system ran at the $\mathrm{2GHz}$ clock frequency and employed 1T1R ReRAM cell structure and 65nm technology node.
The benchmark includes several deconvolutional layers from a set of representative neural networks models including GANs and FCNs.
The details of the benchmark used in our work are summarized in Table~\ref{benchmark}.

The performance of the three designs for each benchmark is provided hereinafter, including latency, energy consumption and area overhead. 
All the results are normalized to that of the zero-padding design. 
For analysis purpose, we present the results by separating the contributions from array and periphery circuitry.
Table~\ref{breakdown} lists the detailed breakdown components and their abbreviations. 


We select the layers from GANs and FCNs as the benchmark in order to evaluate the performance of RED in various deconvolution applications. 
The deconvolution layer in GANs usually has a larger amount of input channels and output channels.
As such, the kernel size is usually large, \textit{e.g.}, $5\times5\times256\times256$ for GAN\_Deconv1. 
In contrast, the kernel size in FCNs is usually much smaller, such as $16\times16\times21\times21$ in voc-fcn8s. 
Such a difference in configuration indicates that in GANs, the array resources could outweigh the peripheral circuitry, while the situation in FCNs is opposite. 
This distinction between GAN and FCN deconvolution is clearly reflected in the evaluation results as we shall present in the following. 


\begin{figure}[b]
\vspace{-12pt}
\begin{center}
\includegraphics[width=0.8\columnwidth]{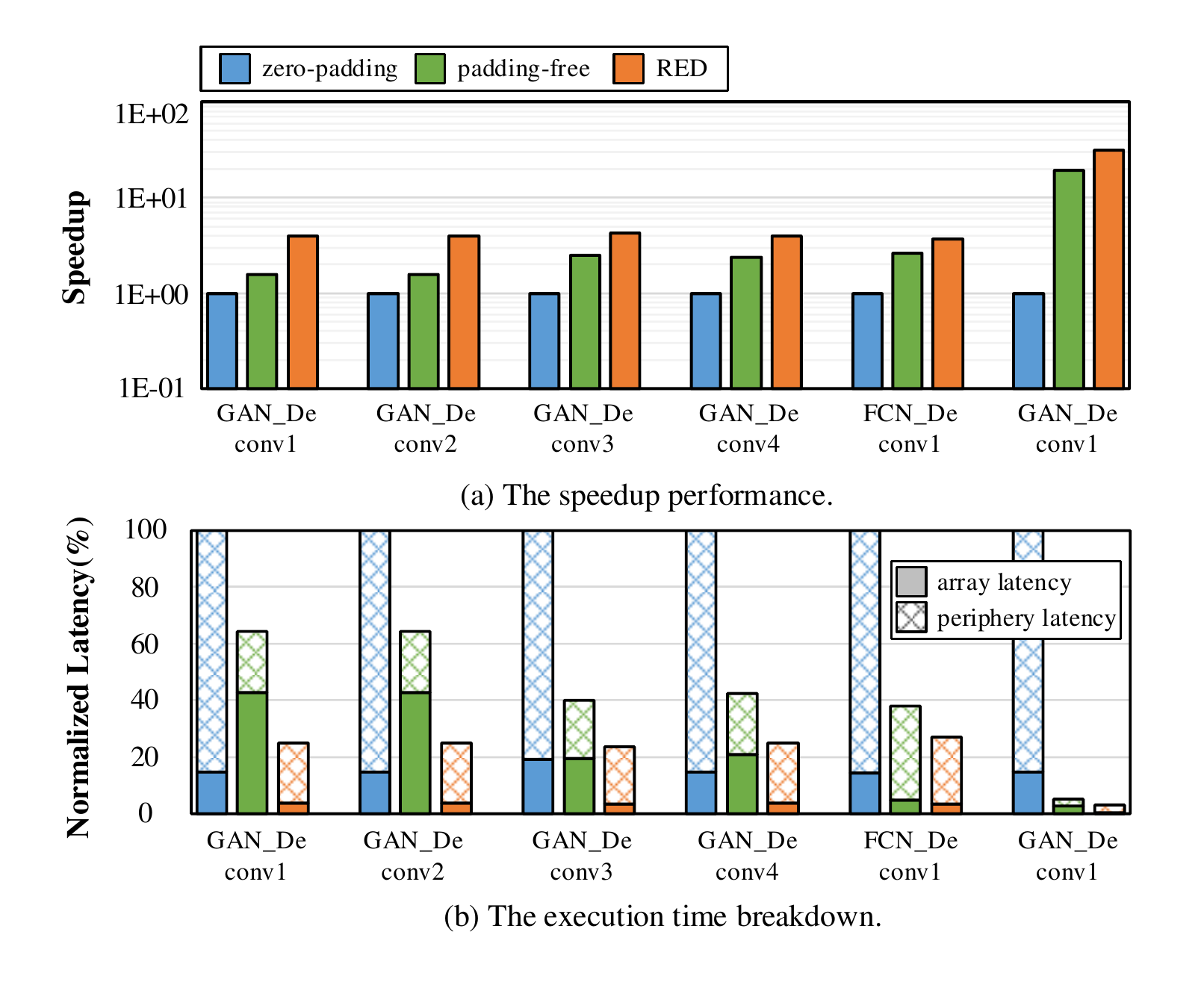}
\vspace{-12pt}
\caption{The latency comparison.}
\label{latency}
\end{center}
\end{figure}

\subsection{Experimental Results \& Analysis}
\subsubsection{Latency}

Fig.~\ref{latency} presents the total and breakdown of latency of the three design implementations obtained from the following calculation:
\begin{equation}
\small
L_{total}=(L_{wd}+L_{bd})_{a}+(L_{dec}+L_{mux}+L_{rc}+L_{sa})_{pp}.
\end{equation}
Fig.~\ref{latency}(a) shows that RED annexes the advantages of both padding-free and zero-padding designs.
It acquires the lowest total latency and achieves highest speedup across all the benchmarks. 
The performance improvement of RED benefits from two aspects: 
1) it eliminates the zero redundancy in input vectors and diminishes the number of cycles; and 2) the size of output vectors is the same as the zero-padding design, hence the two designs have the similar array latency, which is much lower than that of the padding-free design. 
Compared to the zero-padding design, RED achieves $3.69\sim 31.15\times$ speedup.

Fig.~\ref{latency}(b) presents the breakdown of the execution time. 
Compared to the padding-free design, RED reduces the array latency because of the smaller size of output vectors and thus the lower latency caused by wordline driving. 
The padding-free design has longer array latency for its much longer output vector. 
Compared to the zero-padding design, RED arouses 76.9\%$\sim$96.8\% less array and periphery latency. 
The zero-padding design requires $\mathrm{stride^2} \times$ number of cycles compared to the other two designs after adding zero redundancy to input vectors, which induces extensive periphery latency to the computation. 
When $\mathrm{stride=2}$ (such as the GANs and FCN\_Deconv4), the zero-padding design reaches $4\times$ periphery latency compared to the padding-free design and RED. 
Despite the fact that the padding-free design produces more array latency than the zero-padding design and RED in GANs, the zero-padding design still holds $1.55\sim2.62\times$ longer latency than the padding-free design.


\begin{figure}[t]
\begin{center}
\includegraphics[width=0.8\columnwidth]{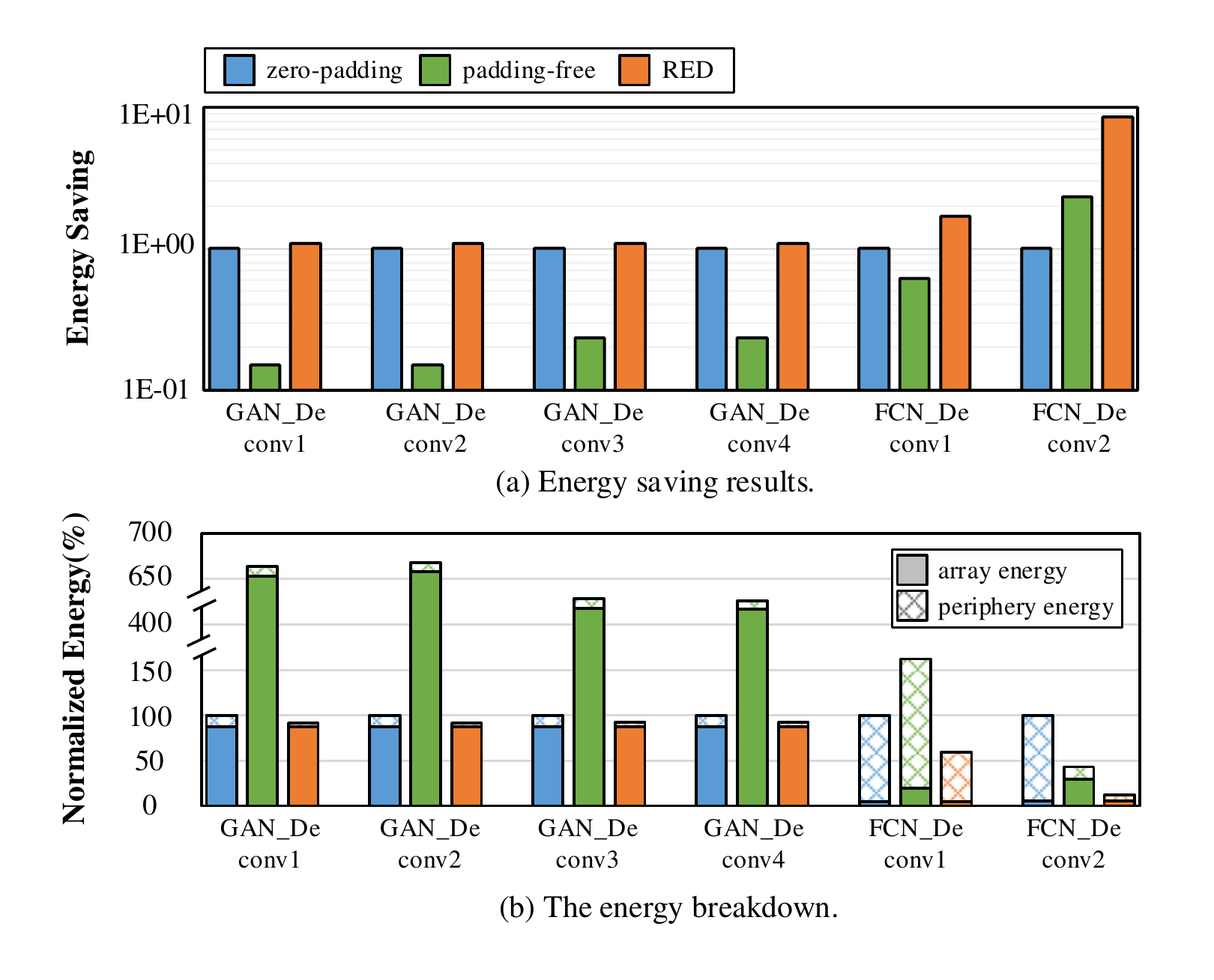}
\vspace{-12pt}
\caption{Then energy comparison.}
\label{energy}
\end{center}
\vspace{-24pt}
\end{figure}

\subsubsection{Energy}
Fig.~\ref{energy} presents the total and breakdown of energy consumption of the padding-free, zero-padding and RED. 
The following equation shows the breakdowns of the energy consumption.
\begin{equation}
\small
E_{total}=(E_{c}+E_{wd}+E_{bd})_{a}+\\(E_{dec}+E_{mux}+E_{rc}+E_{sa})_{pp}.
\end{equation}
Experimental results demonstrate that RED outperforms the other two implementations in the total energy efficiency. 
Owing to the prodigious energy consumption for wordline/bitline driving, the array energy of the padding-free design is conspicuously considerable, which is about $4.48\sim7.53\times$ compared to the other two  designs. 
For this reason, the padding-free design consumes up to $6.68\times$ more energy than the others when implementing GAN where the array contributes more. 

Due to the fact that the total size of the ReRAM crossbar array remains the same, the zero-padding design and RED have the similar array energy. 
The periphery energy of RED is lower than that of the zero-padding design as the input data size of each crossbar is reduced, and thereby decoders consume less energy. 
In total, RED saves $8\%\sim88.36\%$ energy consumption than the zero-padding design. 

\subsubsection{Area}

\begin{figure}[t]
\begin{center}
\includegraphics[width=0.8\columnwidth]{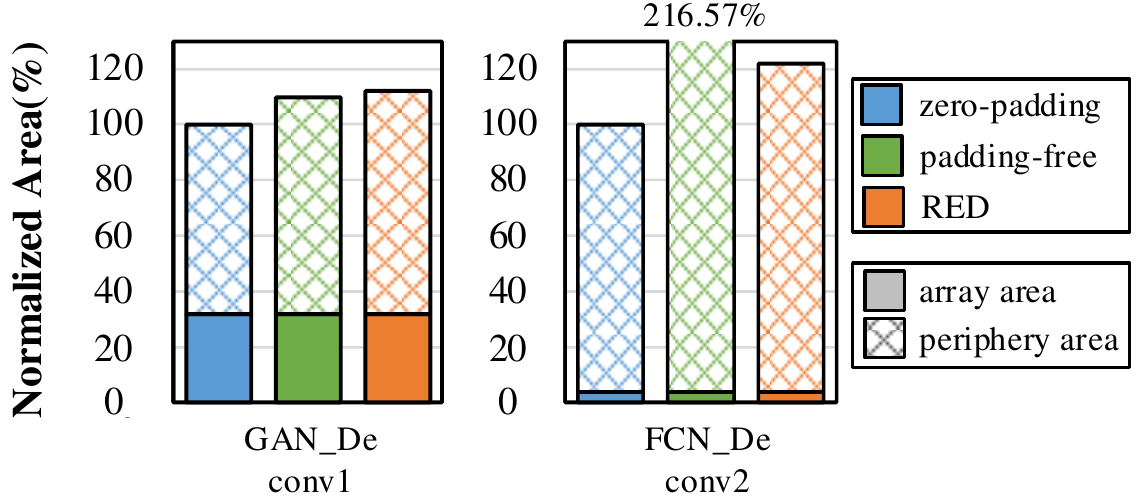}
\vspace{-12pt}
\caption{The area comparison.}
\label{area}
\vspace{-6pt}
\end{center}
\end{figure}

Fig.~\ref{area} shows the breakdown of the area overhead of the three designs. 
For the sake of brevity, we show only a handful of cases. Similar area overhead is observed for all the layers of GANs and FCNs considered in our study. 
Likewise, the area overhead has two parts---array area and periphery area. 
The results demonstrate that three designs incur the same array area because of their identical kernel size. 
The padding-free design procures higher area overhead ($9.79\%$ in GANs and $116.57\%$ in FCNs) in counting of numerous output-related circuits. 
The disparity between the area overhead of the padding-free design and the zero-padding design is remarkable in FCNs. 
The reason is the difference ($K_H\times K_W$ times) in the output sizes of the two designs. More specific, it is $25\times$ in GAN\_Deconv1 but $256\times$ in FCN\_Deconv2. 
Compared with the zero-padding design, the proposed RED introduces $21.41\%$ higher area overhead.
The overhead increases mainly because the pixel-wise mapping method augments output-related periphery circuits by splitting the crossbar apart.

\section{Conclusion}
\label{sec:conclusion}
This work introduces RED, a high-performance and energy-efficient ReRAM-based deconvolution accelerator. 
Through the optimization of the mapping design and data flow, RED eliminates the redundant computations and avoids the overhead of the incremental periphery circuitry. 
Experimental evaluation shows that RED outperforms the existing ReRAM-based accelerators for the common deconvolutional computation algorithms, with up to $31.15\times$ speedup and $88.36\%$ energy consumption reduction. 
\section*{Acknowledgements}
This work was supported by US Department of Energy (DOE) SC0017030. Bing Li acknowledges the National Academy of Sciences (NAS), USA for awarding the NRC research fellowship.

\small
\bibliographystyle{unsrt}
\bibliography{main.bib}
\end{document}